\documentclass[onefignum,onetabnum]{siamart251216}

\usepackage{lipsum}
\usepackage{amsfonts}
\usepackage{graphicx}
\graphicspath{{Figures/}}  
\usepackage{epstopdf}
\usepackage{algorithmic}
\usepackage{tikz}
\usepackage{array}
\usepackage{listings}
\usetikzlibrary{arrows.meta, positioning, shapes.geometric}

\ifpdf
  \DeclareGraphicsExtensions{.eps,.pdf,.png,.jpg}
\else
  \DeclareGraphicsExtensions{.eps}
\fi

\newsiamremark{remark}{Remark}
\newsiamremark{hypothesis}{Hypothesis}
\Crefname{hypothesis}{Hypothesis}{Hypotheses}
\newsiamthm{claim}{Claim}

\headers{Exploring Neural Dynamics with Time Delays}{M. R. Nelson and J. J. Crofts}

\title{Exploring Neural Dynamics with Time Delays: A Numerical and Bifurcation Analysis Approach\thanks{Submitted to the editors 26th October 2026.}}

\author{Martin R. Nelson\thanks{Department of Physical Sciences, Nottingham Trent University, Nottingham, NG11 8NS, United Kingdom 
  (\email{martin.nelson@ntu.ac.uk}, \email{jonathan.crofts@ntu.ac.uk}).}
\and Jonathan J. Crofts\footnotemark[2]}

\usepackage{amsopn}

\ifpdf
\hypersetup{
  pdftitle={Exploring Neural Dynamics with Time Delays: A Numerical and Bifurcation Analysis},
  pdfauthor={M. R. Nelson, and J. J. Crofts}
}
\fi

\begin{document}

\maketitle

\begin{abstract}
 The Wilson--Cowan model is a classic example of a ``neural mass model''—a simplified yet powerful way to describe the activity of large networks of neurons. It is particularly interesting because, even in its relatively simple form, it can produce a surprising variety of behaviours, from steady rhythms to complex oscillations. A key factor influencing these dynamics is the presence of time delays, reflecting the time it takes for signals to travel between neurons. This article offers an accessible introduction to the Wilson--Cowan model, focusing on how these time delays affect its behaviour. We explore the core concepts behind the model, survey the existing literature, and introduce some numerical techniques specifically designed for analysing such delay differential equations. We also provide a practical, hands-on tutorial on using the DDE-BIFTOOL software package, demonstrating its application to the Wilson-Cowan model. Downloadable code allows the reader to reproduce our results and to explore the model's dynamics further, including MATLAB code for all computations, using DDE-BIFTOOL for the continuation, and Python versions of simulation codes.
\end{abstract}



\section{Introduction}
\label{sec:intro}
Computational neuroscience plays a crucial role in bridging the gap between experimental observations of brain activity and our theoretical understanding of how the brain functions \cite{dayan2005theoretical}. It employs mathematical and computational tools to model and simulate neural systems, ranging from individual neurons \cite{burkitt2006review,izhikevich2003simple} to large-scale brain networks \cite{coombes2010large,pathak2022whole,sanz2015mathematical}. This field has a rich history, with early successes like the Hodgkin-Huxley model \cite{hodgkin1952quantitative}, which, through its detailed description of the ionic currents underlying the action potential, earned its creators the Nobel Prize in Physiology or Medicine \cite{schwiening2012brief}. However, even with such detailed models, capturing the full complexity of neural systems remains a challenge. While simplified models can offer valuable insights, incorporating more physiological detail is often crucial for accurately reproducing observed brain dynamics. This is particularly true when considering the influence of factors like the finite speed of signal propagation along axons and the time required for synaptic transmission. These inherent delays in neural communication, often neglected in simpler models, can significantly impact the behaviour of neural circuits \cite{al2021impact,pinder2024network}.

Neural mass models offer a simplified yet powerful approach to understanding the collective dynamics of large populations of neurons \cite{coombes2010large}. These models operate at a mesoscopic scale, averaging the activity of many neurons within a specific brain region to describe the overall behaviour of that population. Rather than simulating the intricate spiking activity of individual neurons, neural mass models focus on capturing the average firing rate and synaptic activity, thus significantly reducing computational complexity. This simplification allows researchers to explore large-scale brain networks and investigate how different brain regions interact. These models are particularly useful for understanding phenomena like oscillations, synchrony, and the impact of various perturbations on brain activity, providing a crucial bridge between microscopic neuronal details and macroscopic brain function measurable with techniques like EEG and MEG \cite{david2003neural,jansen1995electroencephalogram}.  This review focuses on one such crucial physiological detail: time delays and their effect on the dynamics of neural mass models, specifically within the context of the Wilson--Cowan model \cite{wilson1972excitatory}. We will explore how incorporating these delays, which arise from the finite speed of action potential propagation and synaptic transmission, can dramatically alter the predicted behaviour of neural populations, leading to phenomena such as oscillations, synchrony, and even instability and chaos. Understanding the role of time delays is essential for developing more realistic and predictive models of brain function \cite{campbell2007time}.

Within the family of neural mass models, the Wilson--Cowan model stands as a foundational example \cite{wilson1972excitatory}.  This model, developed in the 1970s, describes the interaction between excitatory and inhibitory populations of neurons.  It captures the essential dynamics of neuronal feedback loops, demonstrating how the interplay between excitation and inhibition can give rise to a rich repertoire of behaviours, including stable states, oscillations, and bistability \cite{hoppensteadt2012weakly}.  The Wilson--Cowan model, while relatively simple, provides a valuable framework for understanding fundamental principles of neural population dynamics and serves as a stepping stone for more complex models.  Its relative tractability makes it an ideal system for exploring mathematical techniques applicable to a wider range of neural mass models, including those incorporating time delays, which we will focus on in this review.

We begin this tutorial with a general introduction to delay differential equations (DDEs) and their properties.  We then introduce the Wilson--Cowan model as a key example of a neural mass model, highlighting its structure and biological interpretation.  Following this, we discuss numerical methods suitable for solving DDEs, laying the groundwork for our analysis.  We then delve into basic bifurcation analysis techniques, primarily using the DDE-BIFTOOL software package \cite{engelborghs2002numerical,engelborghs2000dde}, to explore how time delays influence the stability and qualitative behaviour of solutions to the Wilson--Cowan system.  We further examine the emergence of complex dynamics, including chaos, in this model due to the presence of delays.  We provide pointers to further reading for those interested in exploring these topics in greater depth. Finally, we summarise our findings and offer concluding remarks on the importance of considering time delays in neural mass models.

All the code used in this paper (including the listings, the scripts that produce every figure, and Python versions of the simulation codes) is available at \url{https://github.com/martinrnelson/WC_DDE}.


\section{Delay Differential Equations}

Delay differential equation (DDE) models are playing an increasingly prominent role across a range of life science applications \cite{smith2011introduction}. They are particularly crucial in neuroscience \cite{campbell2007time} since neural communication delays necessitate DDEs to model neural dynamics accurately, impacting oscillations, synchrony, and network stability, which is essential for understanding both healthy and diseased brain function.

The focus of this review will be the following class of delay differential equations:
\begin{equation}
 \frac{\mathrm{d}\mathbf{x}}{\mathrm{d}t} = \mathbf{F}(\mathbf{x}(t), \mathbf{x}(t-\tau)), \quad \mathbf{x}(t)\in\mathbb{R}^n, t\geq 0, \label{eqn:dde}
\end{equation}
where $\tau$ is the delay, and $\mathbf{x}(t)\in\mathbb{R}^n$ is a state in configuration space describing the average activity of a neuronal population. In relevant neural dynamics models, the function $\mathbf{F}:\mathbb{R}^{n(d+1)}\to\mathbb{R}^n$ (where $d$ is the number of delays in the system, here equal to one) typically combines a decay term with a sigmoidal input function, providing a simplified yet effective way to model the essential dynamics of neural populations.

The trajectory of a delay differential equation such as that in (\ref{eqn:dde}) is uniquely defined by the associated initial function, $\boldsymbol{\phi}(t)$ such that
\begin{equation}
    \mathbf{x}(t) = \boldsymbol{\phi}(t),\quad t\leq 0.
    \label{eqn:history}
\end{equation}

For clarity, this review focuses on systems with a single, constant time delay. While more complex delay structures are certainly relevant, this simpler case is both mathematically tractable and frequently encountered in modelling applications.  We will briefly touch upon more general delay types and their analysis in Section \ref{sec:advanced}, when discussing future research directions. 

\subsection{Stability Analysis of DDEs}
A crucial first step in understanding the behaviour of a nonlinear system like the one in (\ref{eqn:dde}) is identifying its fixed points (or `equilibria') and determining whether they are stable or unstable. While determining equilibria of DDEs mirrors the approach for ODEs studied in a typical undergraduate dynamics course, stability calculations are more complex in the DDE case due to the presence of the delays.

At a fixed point, the system's state is constant over time. The constant state $\mathbf{x}(t)\equiv\mathbf{x}(t-\tau)\equiv\mathbf{x}^*$ solves (\ref{eqn:dde}) precisely when
\begin{equation}
    \mathbf{F}(\mathbf{x}^*, \mathbf{x}^*) = \mathbf{0}.
    \label{eqn:FP}
\end{equation}
Writing $\mathbf{x}(t) = \mathbf{x}^*+\boldsymbol{\delta}(t)$, where $\boldsymbol{\delta}(t)$ is a small perturbation away from the fixed point $\mathbf{x}^*$, we have
\[
 \dot{\boldsymbol{\delta}}(t) = \dot{\mathbf{x}}(t) = \mathbf{F}\big(\mathbf{x}^*+\boldsymbol{\delta}(t),\, \mathbf{x}^*+\boldsymbol{\delta}(t-\tau)\big),
\]
and we can use Taylor's expansion to obtain
\begin{align}
    \begin{split}
    \mathbf{F}(\mathbf{x}^*+\boldsymbol{\delta}(t), \mathbf{x}^*+\boldsymbol{\delta}(t-\tau)) &= \mathbf{F}(\mathbf{x}^*, \mathbf{x}^*) + D\mathbf{F}(\mathbf{x}^*,\mathbf{x}^*)\boldsymbol{\delta}(t)\\ &+ D\mathbf{F}_\tau(\mathbf{x}^*,\mathbf{x}^*)\boldsymbol{\delta}(t-\tau) + \mathcal{O}(\|\boldsymbol{\delta}\|^2).
    \end{split}
\end{align}
Here, $\mathcal{O}(\|\boldsymbol{\delta}\|^2)$ denotes terms that are quadratic or higher order in the components of $\boldsymbol{\delta}(t)$ and $\boldsymbol{\delta}(t-\tau)$, and $D\mathbf{F}$ and $D\mathbf{F}_\tau$ represent the Jacobian matrices with respect to the state and delayed state variables respectively. Finally, since $\mathbf{x}^*$ represents a fixed point, we know that $\mathbf{F}(\mathbf{x}^*, \mathbf{x}^*)=\mathbf{0}$. Neglecting terms of $\mathcal{O}(\|\boldsymbol{\delta}\|^2)$  then provides the linearised system:
\begin{equation}
    \dot{\boldsymbol{\delta}}(t) = D\mathbf{F}(\mathbf{x}^*,\mathbf{x}^*)\boldsymbol{\delta}(t) + D\mathbf{F}_\tau(\mathbf{x}^*,\mathbf{x}^*)\boldsymbol{\delta}(t-\tau).
    \label{eqn:linear}
\end{equation}

This linear equation describes how small disturbances around the fixed point $\mathbf{x}^*$ grow or decay. To analyse this, we look for solutions of the form $\boldsymbol{\delta}(t) = e^{\lambda t}\mathbf{k}$, where $\lambda$ is a complex number (the growth rate) and $\mathbf{k}$ is a constant vector (the direction of the perturbation), both of which we need to determine. Substituting this into (\ref{eqn:linear}) we obtain
\begin{equation}
    \lambda e^{\lambda t}\mathbf{k} = \left(e^{\lambda t}D\mathbf{F} + e^{\lambda(t-\tau)}D\mathbf{F}_\tau\right)\mathbf{k},
\end{equation}
which, after cancelling the $e^{\lambda t}$ term, rearranges to give
\begin{equation}
    \left(-\lambda I +D\mathbf{F}+e^{-\lambda\tau}D\mathbf{F}_\tau\right)\mathbf{k} = \mathbf{0}.
\end{equation}
For non-zero solutions to exist (\emph{i.e.} $\mathbf{k}\neq \mathbf{0}$), we require
\begin{equation}
    \det\left(-\lambda I +D\mathbf{F}+e^{-\lambda\tau}D\mathbf{F}_\tau\right) = 0.
    \label{eqn:CharEqn}
\end{equation}
Note that (\ref{eqn:CharEqn}) is the \emph{characteristic equation} of the linearised equation in (\ref{eqn:linear}) and that any $\lambda\in\mathbb{C}$ that satisfies (\ref{eqn:CharEqn}) gives rise to a solution of (\ref{eqn:linear}). As for ordinary differential equations, the fixed point $\mathbf{x}^*$ is stable if all the roots $\lambda$ of the characteristic equation have negative real parts. However, the presence of the exponential term in (\ref{eqn:CharEqn}) introduces a significant complication: it leads to infinitely many solutions for $\lambda$ \cite{hale2013introduction}.  This means that standard stability analysis techniques for ordinary differential equations do not directly apply, thus we need specialised methods to determine the stability of DDEs. In practice, this often involves using numerical bifurcation software packages.  In this review, we will focus on DDE-BIFTOOL \cite{engelborghs2002numerical,engelborghs2000dde}.

\subsection{Numerical Methods for DDEs}
Solving delay differential equations in MATLAB is simplified by the built-in {\texttt dde23} function. This versatile solver can handle a variety of DDEs, including those represented by (\ref{eqn:dde}), and is based on a modified Runge--Kutta method, analogous to the {\texttt ode23} solver for ordinary differential equations. (See \cite{shampine2001solving} for more details.) Alongside the MATLAB {\texttt dde23} solver, we also solve (\ref{eqn:dde}) using the Euler map approximation, in which time-discretisation allows us to replace the DDE system with a difference equation:
\begin{align}
\begin{split}
    \mathbf{x}_{k+1} &= \mathbf{x}_k + h\mathbf{F}(\mathbf{x}_k, \mathbf{x}_{k-m}), \quad k\geq 0, \\
    \mathbf{x}_k &= \boldsymbol{\phi}(kh), \quad -m\leq k \leq 0,
\end{split}
\label{eqn:EulerMap}
\end{align}
in which $h = \tau/m$ ($m\in\mathbb{N}$) is the step size and $\mathbf{x}_k$ denotes the numerical approximation of $\mathbf{x}$ at time $t=kh$. We note that, although $\mathbf{x}_{k+1}$ depends only on $\mathbf{x}_k$ and $\mathbf{x}_{k-m}$, continuing the iteration requires all of the intermediate iterates $\mathbf{x}_{k-m},\dots,\mathbf{x}_{k-1}$. For simplicity, we restrict our analysis to step sizes that are integer submultiples of the time delay $\tau$, thus avoiding the need for interpolation. Note that the Euler map offers a computationally efficient approach to the calculation of dynamic observables, such as the Lyapunov exponent (as discussion in Section \ref{sec:chaos}). For such observables, the emphasis is typically on long-term dynamics, where qualitative agreement with the true solution is often prioritised over high levels of numerical accuracy. Additionally, this simplified framework facilitates understanding the fundamental principles of numerically solving DDEs, making it a valuable pedagogical tool.

In \Cref{sec:NumSims} we apply the Euler map to the Wilson--Cowan model and compare it with MATLAB's \texttt{dde23}.


\section{Wilson--Cowan Model}
\label{sec:WCM}

To investigate the impact of time delays on neural dynamics, this work employs the population model of Wilson and Cowan \cite{wilson1972excitatory}. This influential model considers two interacting populations: excitatory and inhibitory neurons. Neural activity within these populations is described by the following system of delay differential equations:
\begin{align}
    \begin{split}
        \frac{\mathrm{d}u}{\mathrm{d}t} &= -u(t) + f\left(au(t-\tau)+bv(t-\tau)+P\right),\\
        \frac{\mathrm{d}v}{\mathrm{d}t} &= -v(t) + f\left(cu(t-\tau)+dv(t-\tau)+Q\right).
    \end{split}
    \label{eqn:WC}
\end{align}
Here, $u$ and $v$ represent the synaptic activity of the two populations, the architecture of the network is given by the weights $a, b, c$ and $d$, while the parameters $P$ and $Q$ are control parameters that represent the basal input to each population. The population firing rate is determined by the sigmoid function
\[
f(z) = \frac{1}{1+e^{-\beta z}}.
\]

Throughout this paper we use two parameter sets, both taken from Coombes and Laing~\cite{coombes2009delays} and listed in \Cref{tab:params}. They are chosen to do different jobs. The first, which we call the \emph{gentle} set, has a shallow sigmoid ($\beta=1$), which renders the bifurcation analysis significantly more computationally tractable. The ``gentle'' parameter set offers a natural setting for a first simulation and a first continuation run, and we use it in \Cref{sec:NumSims} and the first half of \Cref{sec:CompBifn} to introduce the numerical methods. Secondly, the \emph{rich} set, has a steep sigmoid ($\beta=60$), which does render the numerical continuation more difficult, but has the benefit of opening up a much richer array of dynamics and bifurcations. We use this ``rich'' parameter set in the second half of \Cref{sec:CompBifn} and in \Cref{sec:chaos} to show how these higher-order bifurcations are detected and followed. For both sets the delay at which the first oscillation appears can be calculated exactly; see \Cref{sec:WCHopf} and
\Cref{rem:nonsym}.

\begin{table}[tbhp]
  \caption{Parameter sets used throughout this paper, taken from \cite{coombes2009delays}. The penultimate column gives the delay at which the equilibrium first loses stability to an oscillation.}
  \label{tab:params}
  \centering\small
  \setlength{\tabcolsep}{4pt}%
  \begin{tabular}{lrrrrrrrc>{\raggedright\arraybackslash}p{3.3cm}}
    \hline
    Set & $a$ & $b$ & $c$ & $d$ & $P$ & $Q$ & $\beta$ & Hopf delay & Used in\\
    \hline
    Gentle & $10$ & $-10$ & $10$  & $2$  & $-2$  & $-4$  & $1$  & $\tau\approx0.108$
      & \Cref{fig:1}; Listings~\ref{lst:eulermap}--\ref{lst:ddebif}\\
    Rich   & $-6$ & $2.5$ & $2.5$ & $-6$ & $0.2$ & $0.2$ & $60$ & $\tau\approx0.0486$
      & \Crefrange{fig:2}{fig:7};
        Listings~\ref{lst:hopfdelays}--\ref{lst:wcdde23}, \ref{lst:psolsim}, \ref{lst:lyap}\\
    \hline
  \end{tabular}
\end{table}

\subsection{Delay-induced oscillations: an exact calculation}
\label{sec:WCHopf}
\begin{figure}[htbp]
  \centering
  \includegraphics[width=\linewidth]{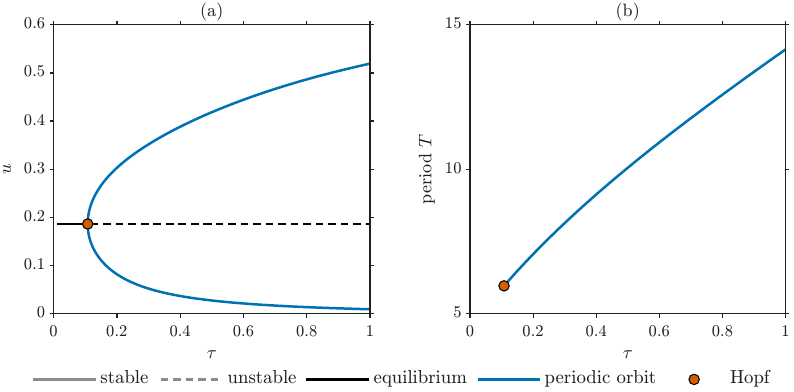}
  \caption{Delay-induced Hopf bifurcation in the Wilson--Cowan model for the ``gentle'' parameters of \Cref{tab:params}. (a) The equilibrium $u^*\approx 0.186$ (black) is stable for small delays and loses stability at a supercritical Hopf bifurcation at $\tau_c\approx 0.108$ (circle), giving rise to a branch of stable periodic orbits; blue curves show the maximum and minimum of $u$ over one period. (b) Period $T$ along the periodic branch. The orbit is born with the finite period $2\pi/\omega\approx 5.97$, where $\omega\approx 1.053$ is the frequency of the critical characteristic roots, and the period then grows approximately linearly with $\tau$.}
  \label{fig:1}
\end{figure}

For the model of (\ref{eqn:WC}), the characteristic equation (\ref{eqn:CharEqn}) can be solved by hand, and doing so tells us exactly when the delay destabilises an equilibrium solution, giving rise to oscillations. We give the calculation for the symmetric case $a=d$, $b=c$, $P=Q$, in which the two populations are identical; the ``rich'' parameter set of \Cref{tab:params} is one example of this. \Cref{rem:nonsym} at the end of this section notes how the same steps apply to the nonsymmetric parameters of \Cref{fig:1}.

At an equilibrium, the delayed and current values coincide, as explained in (\ref{eqn:FP}). So, writing $u(t-\tau)\equiv u(t)\equiv u^*$ (and likewise for $v$), the delay drops out of (\ref{eqn:WC}): equilibria do not depend on $\tau$, only their stability does. Because the populations are identical we can look for a symmetric equilibrium $u^*=v^*$, which must satisfy $u^*=f\big((a+b)u^*+P\big)$. It is simple to prove that such a solution always exists: since $0<f(z)<1$ for all $z$, we know that the function $g(u)=f\big((a+b)u+P\big)-u$ satisfies $g(0)=f(P)>0$ and $g(1)=f(a+b+P)-1<0$, so $g$ must have a root in $(0,1)$ by the intermediate value theorem. In practice we find it numerically; see Listing~\ref{lst:hopfdelays} below.

To compute the stability of the equilibrium, we need to construct the corresponding linearised system of (\ref{eqn:linear}). It is helpful to note that, here, the sigmoid satisfies $f'(z)=\beta f(z)(1-f(z))$, so its slope at the equilibrium is $k=\beta u^*(1-u^*)$. It then follows trivially that, in the notation of (\ref{eqn:linear}), $D\mathbf{F}=-I$ and 
\[
  D\mathbf{F}_\tau=k\begin{pmatrix}a&b\\b&a\end{pmatrix}.
\]
The vectors $(1,1)^T$ and $(1,-1)^T$ are eigenvectors of $D\mathbf{F}_\tau$, with eigenvalues $\eta_s=k(a+b)$ and $\eta_a=k(a-b)$ respectively. Since $D\mathbf{F}=-I$ does not mix these directions, the characteristic equation (\ref{eqn:CharEqn}) factorises into two scalar equations:
\begin{equation}
  \lambda+1=\eta\,e^{-\lambda\tau},\qquad \eta\in\{\eta_s,\eta_a\}.
  \label{eqn:scalarchar}
\end{equation}
The first (for $\eta=\eta_s$) governs \emph{in-phase} (synchronous) perturbations, in which the activities of the two populations move together, and the second (for $\eta=\eta_a$) \emph{anti-phase} perturbations, in which one rises as the other falls.

Our task is now to understand how changes in $\tau$ result in changes in the stability of the equilibrium. That is, does changing $\tau$ result in the solutions of (\ref{eqn:scalarchar}) having a change in the sign of their real part?

As (\ref{eqn:scalarchar}) is a continuous function, the equilibrium can only change stability when a root crosses the imaginary axis. For a real root $\lambda$ to change sign, it would need to cross the point $\lambda=0$, which would require $\eta=1$ in (\ref{eqn:scalarchar}). However, given that this statement is independent of $\tau$, this would not constitute a delay-induced instability. Thus, delay-induced instabilities must arise as a complex conjugate pair of roots together cross the imaginary axis. That is, the instability must occur via a Hopf bifurcation.

At the Hopf bifurcation, we can write $\lambda=\pm i\omega$, and (\ref{eqn:scalarchar}) becomes
\begin{equation}
  1\pm i\omega=\eta\,e^{\mp i \omega\tau},\qquad \eta\in\{\eta_s,\eta_a\}.
  \label{eqn:scalarchar2}
\end{equation}
Computing the modulus of both sides of (\ref{eqn:scalarchar2}) provides the restriction $1+\omega^2=\eta^2$, or equivalently $\omega=\sqrt{\eta^2-1}$, which has real solutions only if $|\eta|>1$. Comparing the arguments of each side of (\ref{eqn:scalarchar2}), and noting that $\arg\left(1+i\omega\right)=\arctan\omega$, we can compute the $\tau$-values at which the Hopf bifurcations occur as
\begin{equation}
  \tau_j=\frac{\arg\eta-\arctan\omega+2\pi j}{\omega},\qquad j=0,1,2,\ldots,
  \label{eqn:HopfDelays}
\end{equation}
where $\arg\eta$ is either zero (if $\eta>1$) or $\pi$ (if $\eta<-1$), and the $2\pi j$ term accounts for the fact that complex arguments are only uniquely defined modulo $2\pi$. Equation (\ref{eqn:HopfDelays}) reveals an infinite sequence of equally spaced Hopf bifurcations, a distance $2\pi/\omega$ apart, which is also the period of the oscillation born at each one. As the delay $\tau$ increases, we obtain an increasingly diverse spectrum of oscillatory solutions that would not otherwise be present in a corresponding ODE model.

Finally, it remains to confirm whether the equilibrium is losing or gaining stability at each Hopf bifurcation. To do so, we differentiate (\ref{eqn:scalarchar}) with respect to $\tau$, which after some rearrangement yields the equation
\begin{equation}
    \label{eqn:dlamdt}\frac{\mathrm{d}\lambda}{\mathrm{d}\tau}=-\frac{\lambda(\lambda+1)}{1+\tau(\lambda+1)},
\end{equation}
which tells us whether $\lambda$ is an increasing or decreasing function of $\tau$. At the Hopf bifurcation, we have $\lambda=\pm i\omega$, and the real-part of (\ref{eqn:dlamdt}) becomes $\omega^2/|1+\tau+i\omega\tau|^2$, which is always positive. Thus, every $\lambda$ crosses the imaginary axis in the direction left to right, corresponding to a loss of stability for the equilibrium configuration.

\begin{lstlisting}[float=tp,caption={M-file \texttt{hopfdelays.m}.},label=lst:hopfdelays]
%HOPFDELAYS  Hopf delays of the symmetric Wilson-Cowan equilibrium
%
% For a = d, b = c, P = Q the linearisation splits into an in-phase and
% an anti-phase mode, each satisfying lambda + 1 = mu*exp(-lambda*tau).

a = -6; b = 2.5; P = 0.2; beta = 60;
f = @(z) 1./(1 + exp(-beta*z));

ustar = fzero(@(u) u - f((a+b)*u + P), [0 1]);  % equilibrium u* = v*
k  = beta*ustar*(1 - ustar);                    % slope f'(z*)
mu = k*[a+b; a-b];                              % in-phase, anti-phase
w  = sqrt(mu.^2 - 1);                           % Hopf frequencies (|mu| > 1)
j  = 0:3;
tau = (angle(mu) - atan(w) + 2*pi*j)./w;        % Hopf delays tau_j, j = 0..3

fprintf('u* = %.5f,  k = %.4f\n', ustar, k)
fprintf('in-phase:   mu = %7.3f, omega = %6.3f, tau_j =%s\n', ...
        mu(1), w(1), sprintf(' %.4f', tau(1,:)))
fprintf('anti-phase: mu = %7.3f, omega = %6.3f, tau_j =%s\n', ...
        mu(2), w(2), sprintf(' %.4f', tau(2,:)))
\end{lstlisting}

Listing~\ref{lst:hopfdelays} turns this calculation into a few lines of Matlab code: it finds $u^*$ with \texttt{fzero} and then evaluates (\ref{eqn:HopfDelays}) for each mode, using the ``rich'' parameters values of \Cref{tab:params}. It reports that the equilibrium loses stability to an anti-phase oscillation at $\tau\approx0.0486$, well before the in-phase mode does at $\tau\approx0.121$. This value gives us a fixed reference point: the continuation computations of \Cref{sec:CompBifn} and the simulations of \Cref{sec:chaos} should both find the oscillation starting at exactly this delay (see \Cref{fig:2,fig:6}). Readers may like to change \texttt{b} and watch the critical delay move; repeating the computation over a range of $b$ produces the Hopf curves of \Cref{fig:5} below.

\begin{remark}[Nonsymmetric parameters]\label{rem:nonsym} Nothing in this argument depends on the symmetry except the explicit form of $\eta$. In general the equilibrium $(u^*,v^*)$ has $u^*\neq v^*$ and is found from both equilibrium equations, and $\eta$ runs over the eigenvalues of $D\mathbf{F}_\tau$, which may be complex. The modulus condition becomes $\omega=\sqrt{|\eta|^2-1}$, while (\ref{eqn:HopfDelays}) and (\ref{eqn:dlamdt}) are unchanged. For the ``gentle'' parameters of \Cref{tab:params} the equilibrium is $(u^*,v^*)\approx(0.1864,0.1338)$, the eigenvalues are $\eta\approx0.874\pm1.160i$, and the first Hopf bifurcation occurs at delay $\tau\approx0.108$, with frequency $\omega\approx1.053$ and period $2\pi/\omega\approx5.97$. This is the bifurcation simulated in \Cref{sec:NumSims} and computed with DDE-BIFTOOL in \Cref{sec:CompBifn}, resulting in \Cref{fig:1}.
\end{remark}

\section{Numerical Simulations of the Wilson--Cowan Model}
\label{sec:NumSims}

\begin{lstlisting}[float=tp,caption={M-file \texttt{eulermap.m}.},label=lst:eulermap]
%EULERMAP  Euler map (2.9) for the Wilson-Cowan model (3.1)
%
% The step is h = tau/m, so the delayed state x(t-tau) is always exactly
% m steps back and no interpolation is needed. Constant history on [-tau,0].

a = 10; b = -10; c = 10; d = 2; P = -2; Q = -4; beta = 1;
f = @(z) 1./(1 + exp(-beta*z));
F = @(x,xd) -x + f([a b; c d]*xd + [P; Q]);  % x = [u; v] at t and t - tau

tau = 0.05; T = 400;                     % also try tau = 0.05 and 0.12
m = 100; h = tau/m; N = round(T/h);     % m steps per delay interval
x = repmat([0.19644; 0.13377], 1, N+m+1);   % columns 1:m+1 = history
for n = m+1:N+m
    x(:,n+1) = x(:,n) + h*F(x(:,n), x(:,n-m));
end
t = ((1:N+m+1) - m - 1)*h;

plot(t, x(1,:), 'b-'), xlabel('t'), ylabel('u(t)')
late = x(1, t > T-50);
amp  = max(late) - min(late)            % oscillation amplitude of u
\end{lstlisting}

The \texttt{eulermap.m} M-file in Listing~\ref{lst:eulermap} applies the Euler map (\ref{eqn:EulerMap}) to the Wilson--Cowan model (\ref{eqn:WC}), using the ``gentle'' parameters of \Cref{tab:params} and the constant history $(u,v)=(u^*+0.01,v^*)$ on $[-\tau,0]$. Because the step is $h=\tau/m$, the delayed state is simply the column \texttt{x(:,n-m)}, so the whole history function is stored in the first $m+1$ columns of \texttt{x}. The function \texttt{F} evaluates the right-hand side of (\ref{eqn:WC}) for a current state and a delayed state, written compactly with the weight matrix $\begin{pmatrix}a&b\\c&d\end{pmatrix}$. Running the code for $\tau=0.05$, $0.12$ and $0.5$ gives oscillation amplitudes \texttt{amp} of essentially zero, $0.0878$ and $0.392$: the equilibrium is stable for the smallest delay and has given way to a stable oscillation for the other two, consistent with the Hopf bifurcation at $\tau\approx0.108$ predicted in \Cref{rem:nonsym}. (We use these parameters here, rather than those of Listing~\ref{lst:hopfdelays}, because the resulting oscillations are large and smooth enough to see clearly.)

\begin{lstlisting}[float=tp,caption={M-file \texttt{wc\_dde23.m}.},label=lst:wcdde23]
%WC_DDE23  Delay-induced Hopf bifurcation in the Wilson-Cowan model
%
% Simulates the model with dde23 for three delays, starting from a small
% perturbation of the equilibrium (which does not depend on tau).

a = 10; b = -10; c = 10; d = 2; P = -2; Q = -4; beta = 1;
f   = @(z) 1./(1 + exp(-beta*z));
rhs = @(t,y,Z) [-y(1) + f(a*Z(1) + b*Z(2) + P);      % Z(:,1) = y(t - tau)
                -y(2) + f(c*Z(1) + d*Z(2) + Q)];
ystar = [0.18644; 0.13377];                          % equilibrium
opts  = ddeset('RelTol', 1e-6, 'AbsTol', 1e-9);

taus = [0.05 0.12 0.5];
for k = 1:3
    sol = dde23(rhs, taus(k), ystar + [0.01; 0], [0 400], opts);
    subplot(3,1,k), plot(sol.x, sol.y(1,:)), ylabel('u')
    late = sol.y(1, sol.x > 350);
    fprintf('tau = %.2f: oscillation amplitude of u = %.4f\n', ...
            taus(k), max(late) - min(late))
end
xlabel('t')
\end{lstlisting}

Listing~\ref{lst:wcdde23} solves the same problem with MATLAB's \texttt{dde23}. The delayed state is passed to the right-hand side as the array \texttt{Z}, whose $k^\mathrm{th}$ column holds $y(t-\tau_k)$; with a single delay, \texttt{Z(1)} and \texttt{Z(2)} are $u(t-\tau)$ and $v(t-\tau)$. The history is the same constant vector as in Listing~\ref{lst:eulermap}. The code reports an oscillation amplitude of essentially zero for $\tau=0.05$, $0.0855$ for $\tau=0.12$ and $0.391$ for $\tau=0.5$: the amplitude grows continuously from zero as $\tau$ crosses the Hopf bifurcation, the signature of a supercritical Hopf bifurcation. The Euler map gave $0.0878$ and $0.392$. The difference is largest close to the bifurcation, where the amplitude is most sensitive to the parameters, and it shrinks as \texttt{m} is increased, since the Euler map is first-order accurate in $h$.

\section{Computational Bifurcation Analysis}
\label{sec:CompBifn}

\begin{lstlisting}[float=tp,caption={M-file \texttt{ddebiftool\_hopf.m}.},label=lst:ddebif]
%DDEBIFTOOL_HOPF  Equilibria, Hopf point and periodic orbits as tau varies
%
% Requires DDE-BIFTOOL (v3.x) on the MATLAB path.
% Parameters p = [a b c d P Q beta tau]; the delay is parameter 8.

par = [10 -10 10 2 -2 -4 1 0.01];  itau = 8;
f   = @(z,beta) 1./(1 + exp(-beta*z));
rhs = @(x,p) [-x(1,1) + f(p(1)*x(1,2) + p(2)*x(2,2) + p(5), p(7));
              -x(2,1) + f(p(3)*x(1,2) + p(4)*x(2,2) + p(6), p(7))];
funcs = set_funcs('sys_rhs', rhs, 'sys_tau', @() itau);  % x(:,2) = x(t-tau)

% 1. Branch of equilibria in tau, with stability
stst = SetupStst(funcs, 'x', [0.18644; 0.13377], 'parameter', par, ...
    'contpar', itau, 'step', 0.002, 'max_bound', [itau 0.5], ...
    'max_step', [itau 0.005]);
stst = br_contn(funcs, stst, 100);
[nunst, ~, ~, stst.point] = GetStability(stst, 'funcs', funcs);

% 2. Locate the Hopf point, where two roots cross the imaginary axis
ih   = find(diff(nunst) == 2, 1);
hopf = p_tohopf(funcs, stst.point(ih));
mth  = df_mthod(funcs, 'hopf');
hopf = p_correc(funcs, hopf, itau, [], mth.point);
fprintf('Hopf at tau = %.5f, omega = %.4f\n', hopf.parameter(itau), hopf.omega)

% 3. Branch of periodic orbits from the Hopf point, with Floquet multipliers
po = SetupPsol(funcs, stst, ih, 'degree', 4, 'intervals', 30, ...
    'max_bound', [itau 1], 'max_step', [itau 0.01]);
po = br_contn(funcs, po, 150);
po = br_stabl(funcs, po, 0, 1);
[nunst_po, ~, ~, po.point] = GetStability(po, 'funcs', funcs, ...
                                          'exclude_trivial', true);
tau = arrayfun(@(p) p.parameter(itau), po.point);
amp = arrayfun(@(p) max(p.profile(1,:)) - min(p.profile(1,:)), po.point);
plot(tau, amp, 'b.-'), xlabel('\tau'), ylabel('amplitude of u')
fprintf('unstable periodic orbits found: %d\n', nnz(nunst_po > 0))
\end{lstlisting}

We now turn to the Matlab package DDE-BIFTOOL, for the numerical computation of bifurcation diagrams for our DDE system. Many readers will be familiar with the underlying methodology from similar numerical exploration of ODE systems; the software applies numerical predictor--corrector algorithms to trace out branches of bifurcation curves in a potentially high-dimensional parameter/delay-space. (For a tutorial that considers related numerical bifurcation analysis in the undelayed case, the reader is directed to \cite{gandy2022analyzing}.)

Listing~\ref{lst:ddebif} presents a simple initial exploration of the delay-induced Hopf bifurcations, and resulting periodic orbits, that the Wilson--Cowan model displays for the ``gentle'' parameters of \Cref{tab:params} -- a computationally more-tractable problem than the corresponding ``rich'' set. This code produces the plots shown in \Cref{fig:1}, which illustrates the first Hopf bifurcation for the gentle parameter set, and the growth of the corresponding orbit. The initial prescription of the model involves defining a vector of parameters (\texttt{par} or \texttt{p}), which also includes the delay $\tau$. The right-hand side (\texttt{rhs}) of the system is written in terms of the array \texttt{x}, whose first and second columns contain the dependent variables ($u$, $v$) evaluated at the current state and the delayed state respectively. We call the function \texttt{set\_funcs} to construct a structure containing the above details, and also specify that our delay lies in the eighth position in our parameter list via the flag \texttt{sys\_tau}. We then perform the following three steps. In Step~1, we initially provide the coordinates of an equilibrium state (determined either analytically as described in e.g. Listing~\ref{lst:hopfdelays}, or numerically as the long-term solution to an Euler map or \texttt{dde23} computation -- see Listings~\ref{lst:eulermap} and \ref{lst:wcdde23}), and track this equilibrium and its stability as $\tau$ increases via the branch continuation (\texttt{br\_contn}) routine. For the stability calculation, the variable \texttt{nunst} records the number of eigenvalues that have positive real part.

Step~2 detects Hopf bifurcation points by finding neighbouring points between which the \texttt{nunst} count jumps by two and then performing a correction step to determine the Hopf point accurately; the code reports the first Hopf bifurcation at $\tau = 0.10776$, corresponding with an orbit of frequency $\omega = 1.0532$. Step~3 starts a branch of periodic orbits from the Hopf point, and tracks the evolution of these orbits as the delay increases. As the corresponding branch is computed, the call to the \texttt{GetStability} method computes the stability of the orbit at each point on the branch, this time reporting the Floquet multipliers for the orbit. (Floquet multipliers describe the stability of an orbit in a manner analogous to how the eigenvalues of the Jacobian matrix describe the stability of an equilibrium. A periodic orbit is stable if all of its Floquet multipliers satisfy $|\mu_j|\leq1$.)   As \Cref{fig:1} illustrates, the amplitude of the orbit grows from zero at $\tau_c$ like $\sqrt{\tau-\tau_c}$, as expected at a supercritical Hopf bifurcation. These are the data behind \Cref{fig:1}; code to produce the full figure, which adds the equilibrium branch and marks stability, is available online.

\begin{figure}[htbp]
  \centering
  \includegraphics[width=\linewidth]{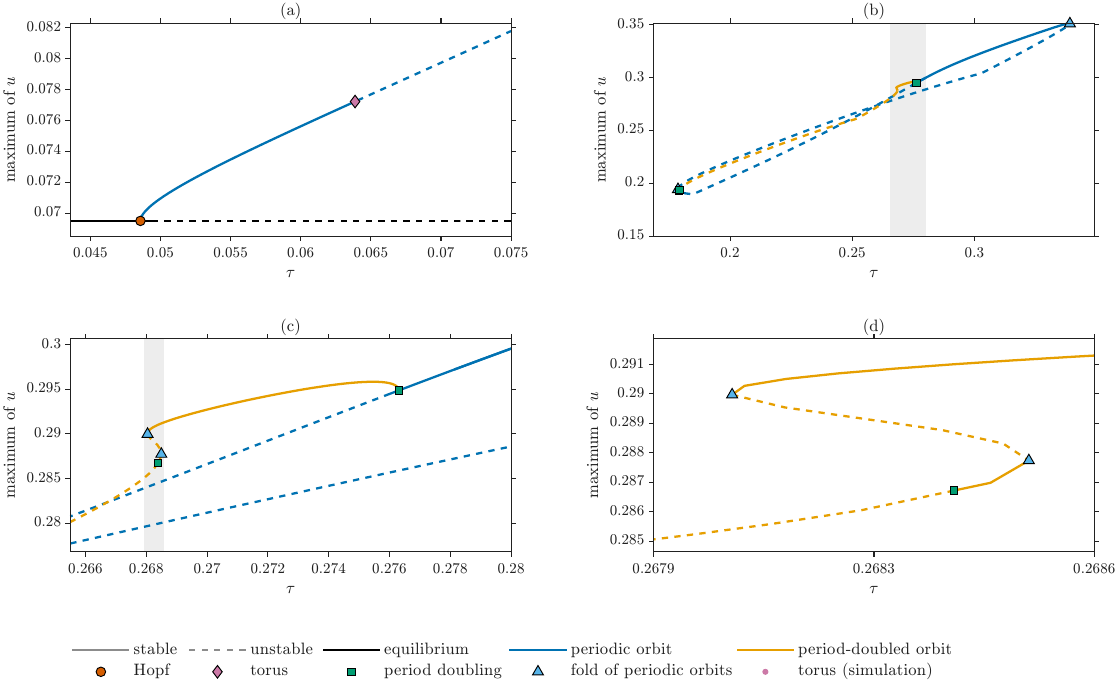}
  \caption{Secondary bifurcations of periodic orbits as $\tau$ is varied, for the ``rich'' parameters of \Cref{tab:params}. Solid/dashed curves represent stable/unstable branches respectively. (a) The equilibrium $u^*\approx 0.0695$ (black) undergoes a Hopf bifurcation at $\tau_\mathrm{H}\approx 0.0486$ (circle), creating an anti-phase periodic orbit, which loses stability at a torus bifurcation at $\tau_\mathrm{T}\approx 0.064$ (diamond). (b) The asymmetric periodic orbits form an isola (blue), a closed branch bounded by folds of periodic orbits at $\tau\approx0.179$ and $\tau\approx0.339$ (triangles) and carrying two period doublings, at $\tau_\mathrm{PD}\approx0.179$ and $\tau_\mathrm{PD}\approx0.276$ (squares). The period-doubled orbits (orange) form a bubble joining these two period doublings. The mirror-image isola, obtained by exchanging $u$ and $v$, is not shown. (c) Enlargement of the shaded region in (b). As $\tau$ decreases, the asymmetric orbit, with two maxima per period, undergoes a supercritical period doubling at $\tau_\mathrm{PD}$, and each maximum splits into two along the stable period-doubled branch. This branch loses stability at a fold at $\tau\approx 0.2680$, continues as an unstable orbit to a second fold at $\tau\approx 0.2685$, where it regains stability, and then undergoes a further period doubling at $\tau_\mathrm{PD}\approx 0.2684$. Two stable period-doubled orbits therefore coexist in the narrow window $0.2684<\tau<0.2685$. (d) Enlargement of the shaded region in (c), showing one of the local maxima of the period-doubled orbit.}
  \label{fig:2}
\end{figure}

We now turn our attention to the ``rich'' parameter set of \Cref{tab:params}. We apply the same steps as described above initially, although it transpires that the steeper sigmoid that results from choosing $\beta=60$ requires some numerical modifications to the code of Listing~\ref{lst:ddebif}; we relegate the details to the supporting codes available online. \Cref{fig:2}(a) illustrates the first Hopf bifurcation at $\tau\simeq0.048$, and the corresponding branch of periodic orbits. Unlike for the ``gentle'' parameter set above, we observe that the resulting stable periodic orbit is soon destabilised at a torus bifurcation, which we identify in DDE-BIFTOOL by observing that a pair of complex conjugate Floquet multipliers move outside of the unit circle on the complex plane (see \Cref{fig:3}). We explore the dynamics beyond the torus bifurcation in more detail in \Cref{sec:chaos} below.

\begin{figure}[htbp]
  \centering
  \includegraphics[width=0.75\linewidth]{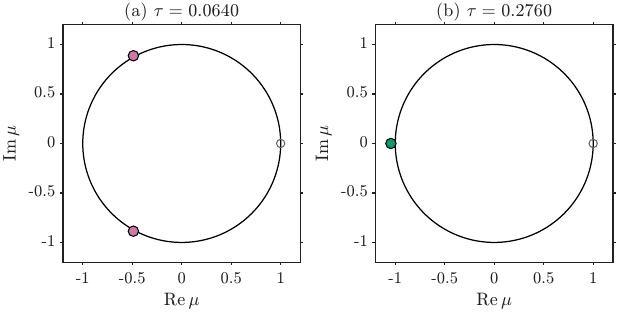}
  \caption{Floquet multipliers $\mu$ of the periodic orbits in \Cref{fig:2}, just past each bifurcation; filled symbols lie outside the unit circle and the open symbol at $\mu=1$ is the trivial multiplier. (a) Anti-phase orbit at $\tau=0.0640$: a complex pair has crossed the unit circle near $\mu\approx -0.5\pm 0.87i$ (torus bifurcation), with argument close to $\pm 2\pi/3$. (b) Asymmetric orbit at $\tau=0.2760$: a real multiplier has passed through $\mu=-1$ (period doubling).}
  \label{fig:3}
\end{figure}

For the ``rich'' parameters, we additionally observe that not every periodic orbit of interest lies on a branch that starts at a Hopf point. The asymmetric orbit of \Cref{fig:2}(b) is an example of this. This renders the continuation a little more difficult as we have no equilibrium to start from in this case. In Listing~\ref{lst:psolsim}, we show how to contruct a branch of orbits starting with an orbit obtained from a \texttt{dde23} simulation at $\tau=0.28$. One period is extracted between successive occurrences of the largest maximum of $u$, and a corresponding DDE-BIFTOOL structure is assembled manually from this data, and corrected by Newton's method (\texttt{p\_correc}) to obtain sufficient accuracy to enable continuation of the branch over increasing $\tau$. Having run the code of Listing~\ref{lst:psolsim}, we perform the continuation exactly as before -- see full code online.

\begin{lstlisting}[float=tp,caption={M-file \texttt{psol\_from\_sim.m}.},label=lst:psolsim]
%PSOL_FROM_SIM  Start a periodic-orbit computation from a simulation
%
% Requires DDE-BIFTOOL and funcs, itau from ddebiftool_hopf.m.
% Uses the beta = 60 parameters, where this orbit is not on the branch
% from the Hopf point, so it must be found another way.

par = [-6 2.5 2.5 -6 0.2 0.2 60 0.28];
fz  = @(z) 1./(1 + exp(-60*z));
dde = @(t,y,Z) [-y(1) + fz(-6*Z(1) + 2.5*Z(2) + 0.2);
                -y(2) + fz(2.5*Z(1) - 6*Z(2) + 0.2)];
sol = dde23(dde, par(itau), [0.3; 0.1], [0 300], ...
            ddeset('RelTol', 1e-6, 'AbsTol', 1e-9));

% one period = time between successive occurrences of the largest maximum
t = linspace(250, 300, 50001);  u = deval(sol, t, 1);
ip  = find(u(2:end-1) > u(1:end-2) & u(2:end-1) >= u(3:end)) + 1;
big = ip(u(ip) > max(u(ip)) - 1e-3);
T0  = t(big(2)) - t(big(1));

deg = 4; ntst = 120; mesh = linspace(0, 1, ntst*deg + 1);
po = struct('kind', 'psol', 'parameter', par, 'mesh', mesh, 'degree', deg, ...
            'profile', deval(sol, t(big(1)) + mesh*T0), 'period', T0);
mth = df_mthod(funcs, 'psol');
po  = p_correc(funcs, po, [], [], mth.point);       % Newton, tau fixed
po  = p_remesh(po, deg, ntst);                      % adapt mesh to profile
[po, suc] = p_correc(funcs, po, [], [], mth.point);

po.stability = p_stabil(funcs, po, mth.stability);
mu = po.stability.mu;  [~, i1] = min(abs(mu - 1));
fprintf('converged = %d, period = %.4f\n', suc, po.period)
fprintf('trivial multiplier = %.4f, most negative = %.4f\n', ...
        real(mu(i1)), min(real(mu)))
\end{lstlisting}

\begin{figure}[htbp]
  \centering
  \includegraphics[width=\linewidth]{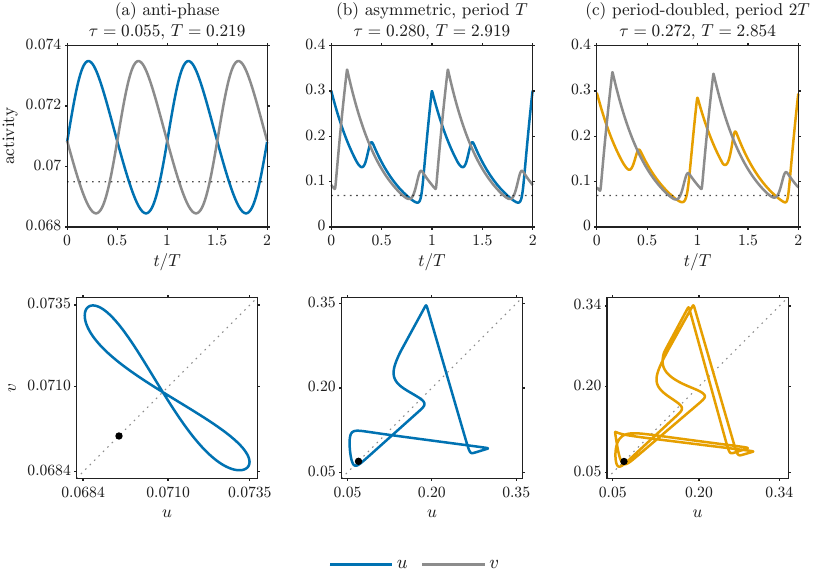}
  \caption{Representative periodic orbits for the ``rich'' parameters of \Cref{tab:params}. Top row: $u(t)$ (coloured) and $v(t)$ (grey) over two periods of the period-$T$ orbit, with the equilibrium value $u^*$ dotted. Bottom row: projections onto the $(u,v)$ plane, with the symmetry line $u=v$ (dotted) and the equilibrium (dot). (a) Anti-phase orbit at $\tau=0.055$ ($T=0.219$): one maximum per period and $v(t)=u(t+T/2)$, so the projection is symmetric about $u=v$. (b) Asymmetric orbit at $\tau=0.280$ ($T=2.919$): two maxima of different heights per period and no reflection symmetry.  (c) Period-doubled orbit at $\tau=0.272$ (period $2T$, $T=2.854$): successive maxima alternate in height, and the projection winds twice around a closed curve.}
  \label{fig:4}
\end{figure}

Continuing this orbit in both directions shows where it comes from. It lies on an \emph{isola}: a closed curve of periodic orbits, bounded by two folds, at $\tau\approx0.179$ and $\tau\approx0.339$, at which pairs of orbits are created and destroyed (\Cref{fig:2}(b)). The orbits on the isola are neither in-phase nor anti-phase (see \Cref{fig:4}(b)) and never meet the corresponding branches that arise from the equilibrium. (We note that due to the $u\leftrightarrow v$ symmetry in our model, there is additionally a mirror-image isola, which we omit in \Cref{fig:2}(b) for clarity.) We see that the orbit is stable in the approximate range $\tau\in[0.276,0.335]$. Moving right to left in \Cref{fig:2}(b), the orbit is destabilised via a period doubling bifurcation at $\tau_{PD}\approx0.276$, at which point a new orbit of twice the period is born, as shown in \Cref{fig:4}(c). (The period doubling bifurcation is identifiable by noting that a single real Floquet multiplier moves through the point $\mu=-1$, as shown in \Cref{fig:3}(b).) The period-doubled orbit branch itself displays various folds and further period doublings (all of which can be detected via DDE-BIFTOOL, code online) before itself being anhiliated through collision with the original isola branch at another period doubling point thereon. (See the enhanced plots of \Cref{fig:2}(c,d) for further detail.) While we could, in principle, continue to trace further branches of orbits arising at each of these additional period doubling points, we omit these here.

\begin{figure}[htbp]
  \centering
  \includegraphics[width=0.7\linewidth]{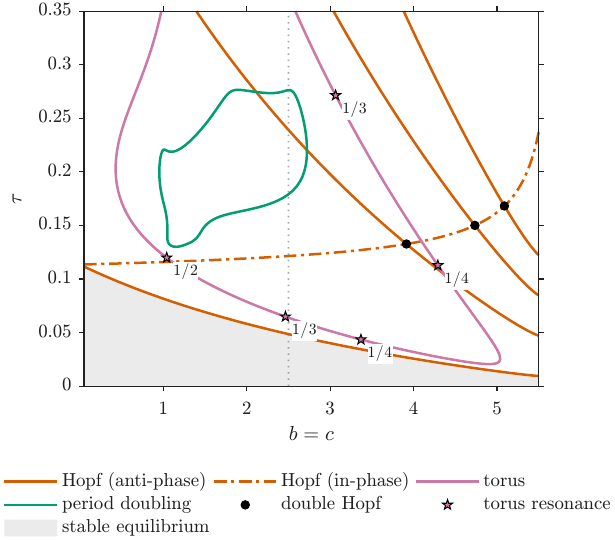}
  \caption{Two-parameter bifurcation diagram in the plane of the cross-coupling $b=c$ and the delay $\tau$, with the remaining parameters as in \Cref{fig:2}; the dotted vertical line marks the one-parameter slice $b=2.5$ studied there. Hopf curves of the equilibrium are computed analytically for the anti-phase (solid) and in-phase (dash-dot) modes; the equilibrium is stable in the shaded region, and the curves intersect at double Hopf points (dots). The torus curve (purple) and the period-doubling curve (green) were obtained by numerical continuation with DDE-BIFTOOL. Stars mark points on the torus curve where the rotation number is rational (i.e. points of mode locking).}
  \label{fig:5}
\end{figure}

It is often of interest to monitor how the various bifurcations of e.g. \Cref{fig:2} move as we vary one or more of the model's other parameters. This can be achieved in DDE-BIFTOOL by setting up corresponding branches of Hopf, torus, or period doubling points, and tracking these using extensions of the continuation tools described above. (For full details, see the supporting code online.) \Cref{fig:5} shows how the bifurcations found for $b=2.5$ in \Cref{fig:2} move as $b$ is varied (keeping $b=c$ here, to preserve symmetry in parameters). The equilibrium is stable in the shaded region, below the first anti-phase Hopf curve. (As \Cref{fig:5} shows, additional branches of Hopf bifurcations also exist for larger values of $\tau$, on which the equilibrium becomes increasingly unstable, and additional unstable orbits are created.) The anti-phase orbit born at the first Hopf curve loses stability on the torus curve (purple), at which point solutions on a surrounding invariant torus are born. At each point of the torus bifurcation curve, a pair of complex conjugate Floquet multipliers lies on the unit circle, with coordinates $\mu=e^{\pm2\pi i\theta}$, where $\theta$ is the \emph{rotation number}: roughly, this represents the fraction of a turn that a small perturbation of the orbit rotates in one period. Crossing the torus curve at a generic point where $\theta$ is irrational results in solutions on the torus winding around it without ever closing up, which is the quasi-periodic motion seen just after $\tau_T$ in \Cref{fig:6}(a) and in \Cref{fig:7}(b). Occasionally on the torus curve, as demarked by stars in \Cref{fig:5}, $\theta=p/q$ is rational, and solutions near the torus can instead `mode lock' onto a periodic orbit that closes after $q$ periods -- see \Cref{fig:7}(c) for an example of this. Additionally, \Cref{fig:5} shows that the Hopf branch corresponding to in-phase solutions can cross branches of Hopfs corresponding to anti-phase phase solutions, giving rise to ``double Hopf'' bifurcations -- co-dimension 2 bifurcations that are known to give rise to additional branches of higher-order bifurcations and corresponding intricate solutions. We direct the reader to e.g. \cite{pinder2023bifurcations} for further details on this topic. 

So far, we have established the following. A unqiue equilibrium in the undelayed model displays an infinite sequence of Hopf bifurcations as we increase the delay. (We have explored the first of these in \Cref{fig:2}(a).) The resulting orbits can then be destabilised by torus bifurcations. Additionally, we have seen isolas of periodic orbits that display period doubling bifurcations and folds of orbits. In essence, the theme here is that as the delay is increased, the dynamics get progressively more intricate. Indeed, traversing an intricate sequence of torus or period doubling bifurcations is a hallmark route to chaotic dynamics, which we explore further below.

\section{Chaos in the Wilson--Cowan Model}
\label{sec:chaos}

We examine chaotic solutions in the Wilson--Cowan model by computing Lyapunov exponents as the delay is increased. Lyapunov exponents are essentially a measure of the extent to which trajectories starting at neighbouring points in phase-space diverge from one another, with positive Lyapunov exponents indicating chaotic behaviour. Listing~\ref{lst:lyap} estimates the largest Lyapunov exponent (LLE, denoted $\lambda_{\max}$) by following two nearby solutions with the Euler map, written as in Listing~\ref{lst:eulermap} but this time for the ``rich'' parameter set. Since the state of a DDE at time $t$ is the whole function on $[t-\tau,t]$, the distance between the two solutions is measured over all $m+1$ stored values, not just the current one. The last $m+1$ states are kept in a circular buffer, so the delayed state is always the oldest column and is overwritten as soon as it has been used. (For further information on Lyapunov exponents and their computation in the DDE case, we direct the reader to \cite{farmer1982chaotic} and \cite{wolf1985determining}.)

\begin{figure}[htbp]
  \centering
  \includegraphics[width=0.85\linewidth]{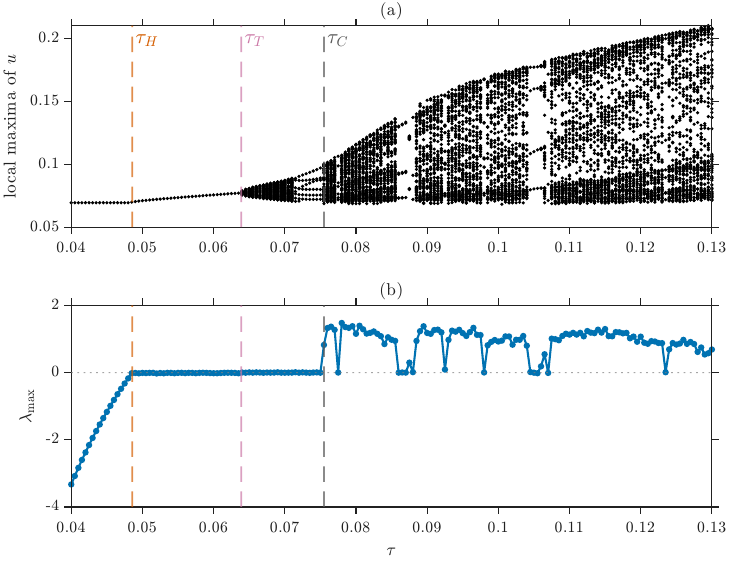}
  \caption{Route to chaos through torus breakdown as the delay is increased, for the ``rich'' parameters of \Cref{tab:params}. (a) Orbit diagram: local maxima of $u(t)$ after transients, computed by direct simulation from a single initial history for each $\tau$.
  (b) Largest Lyapunov exponent $\lambda_{\max}$. The equilibrium loses stability at the Hopf bifurcation $\tau_\mathrm{H}\approx 0.0486$, and the resulting  anti-phase orbit undergoes a torus bifurcation at $\tau_\mathrm{T}\approx 0.064$,  after which the maxima fill an interval and $\lambda_{\max}\approx 0$  (quasi-periodic motion). Windows of mode locking follow, and from  $\tau_\mathrm{C}\approx 0.0755$ the exponent becomes clearly positive (chaos).}
  \label{fig:6}
\end{figure}

\begin{figure}[htbp]
  \centering
  \includegraphics[width=\linewidth]{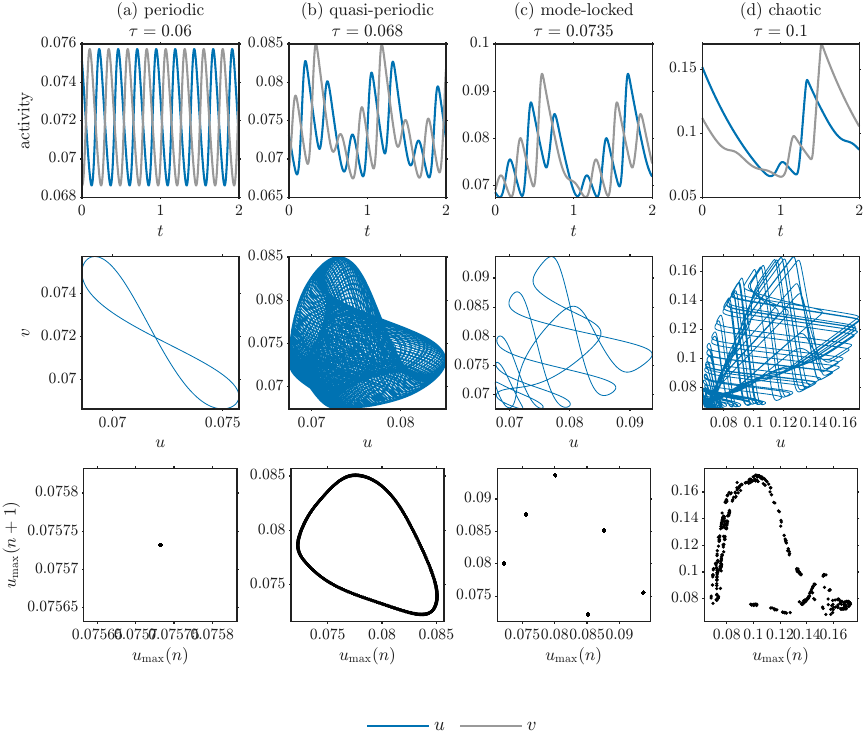}
  \caption{Solutions along the route to chaos in \Cref{fig:6}. Top row: $u(t)$ (blue) and $v(t)$ (grey) over a short time window. Middle row: projection onto the $(u,v)$ plane. Bottom row: return map of successive maxima of $u$. (a) Periodic anti-phase orbit, $\tau=0.060$: a single fixed point of the return map. (b) Quasi-periodic solution after the
  torus bifurcation, $\tau=0.068$: the return map traces a closed invariant curve. (c) Mode-locked solution, $\tau=0.0735$: the invariant curve is replaced by a periodic orbit, seen as a finite set of points. (d) Chaotic solution, $\tau=0.1$: the invariant curve has broken up and the return map fills a folded set.}
  \label{fig:7}
\end{figure}

\Cref{fig:6} illustrates the computed LLE over a range of $\tau$ values, alongside a scatter plot depicting the local maxima that corresponding orbits exhibit. A selection of the corresponding trajectories is shown in \Cref{fig:7}. In \Cref{fig:6}, the Hopf and torus points of \Cref{fig:2}(a) are demarked by $\tau=\tau_\mathrm{H}$ and $\tau=\tau_\mathrm{T}$ respectively. For $\tau<\tau_\mathrm{T}$, the LLE is negative and solutions have a single local maximum as expected -- solutions in this range are either a stable equilibrium ($\tau<\tau_\mathrm{H}$) or a simple orbit ($\tau\in[\tau_\mathrm{H},\tau_\mathrm{T}]$) as depicted in \Cref{fig:4}(a) and \Cref{fig:7}(a). As we cross the first torus bifurcation at $\tau_\mathrm{T}$, and indeed the sequence of torus bifurcations that follow, solutions with more complex shapes emerge, including quasiperiodic and mode-locked states as shown in \Cref{fig:7}(b,c). For these solutions in $\tau\in[\tau_\mathrm{T},\tau_\mathrm{C}]$, where $\tau_\mathrm{C}\simeq0.0755$, the LLE is negative but very close to zero; the solutions are not yet chaotic, but are starting to become heavily unpredictable. As we cross the point $\tau=\tau_\mathrm{C}$, the LLE becomes positive, and chaotic dynamics emerge, resulting in convergence to complex attractors such as that shown in \Cref{fig:7}(d).

\begin{lstlisting}[float=tp,caption={M-file \texttt{lyapdde.m}.},label=lst:lyap]
%LYAPDDE  Largest Lyapunov exponent of the Wilson-Cowan DDE
%
% Two nearby solutions are advanced with the Euler map. The state of a
% DDE is the whole history on [t-tau, t], so the separation is measured
% over all m+1 stored values and renormalised once per delay interval.

a = -6; b = 2.5; P = 0.2; beta = 60; tau = 0.05;   % also try tau = 0.068
f = @(z) 1./(1 + exp(-beta*z));
F = @(x,xd) -x + f([a b; b a]*xd + P);            % rhs for x(t), x(t-tau)
m = 200; h = tau/m;

X = repmat([0.3; 0.1], 1, m+1);   % circular buffer of the last m+1 states
i = m+1;                          % column holding the current state
for n = 1:round(100/h)            % discard a transient of length 100
    j = mod(i, m+1) + 1;          % oldest column = state at t - tau
    X(:,j) = X(:,i) + h*F(X(:,i), X(:,j));
    i = j;
end

d0 = 1e-8;  Y = X + d0/sqrt(numel(X));   % perturbed copy, distance d0
S = 0;  N = round(200/h);
for n = 1:N
    j = mod(i, m+1) + 1;
    X(:,j) = X(:,i) + h*F(X(:,i), X(:,j));
    Y(:,j) = Y(:,i) + h*F(Y(:,i), Y(:,j));
    i = j;
    if mod(n, m) == 0             % once per delay interval
        D = Y - X;  s = norm(D(:));
        S = S + log(s/d0);  Y = X + (d0/s)*D;
    end
end
lambda = S/(N*h)
\end{lstlisting}

\section{Advanced Topics and Future Directions}
\label{sec:advanced}

The model in this review is deliberately minimal: two populations, one fixed delay and no randomness. Each of these simplifications can be relaxed. Below we describe four potential extensions, briefly outlining any necessary modifications to the approach and anticipated results.

\subsection{Networks of delayed neural masses}
\label{sec:networks}
Models of the whole brain often link many neural masses, one for each brain region, using a map of the connections between regions obtained from brain imaging. Signals between distant regions take longer to arrive, so each connection has its own delay \cite{breakspear2017dynamic,sanz2015mathematical}.

Each brain region is represented by a copy of the model (\ref{eqn:WC}), with its own excitatory and inhibitory populations, and the regions are coupled through long-range connections, typically between excitatory populations. With $N$ regions, the model then becomes
\begin{align}
  \begin{split}
  \frac{\mathrm{d}u_i}{\mathrm{d}t} &= -u_i(t)+ f\Big(a u_i(t-\tau) + b v_i(t-\tau) + \varepsilon\sum_{j=1}^{N} w_{ij}\,u_j(t-\tau_{ij}) + P\Big),\\
  \frac{\mathrm{d}v_i}{\mathrm{d}t} &= -v_i(t)+ f\big(c u_i(t-\tau) + d v_i(t-\tau) + Q\big),
  \end{split}
  \label{eqn:network}
\end{align}
for $i=1,\ldots,N$. Here $u_i$ and $v_i$ are the excitatory and inhibitory activities in region $i$, and $a$, $b$, $c$, $d$ and $\tau$ describe the coupling within a region, exactly as in (\ref{eqn:WC}). The weight $w_{ij}$ is the strength of the connection from region $j$ to region $i$, taken from imaging data, $\varepsilon$ sets the overall strength of the long-range coupling, and $\tau_{ij}$ is the delay along each connection. With a single region ($N=1$, $\varepsilon=0$) we recover (\ref{eqn:WC}). When all the delays are equal, the change to the code is small. Listing~\ref{lst:eulermap} can be updated to prescribe the right-hand side of (\ref{eqn:network}) as follows: 
\begin{lstlisting}[frame=none]
I = eye(N);  M = [a*I + epsilon*W, b*I;  c*I, d*I];
F = @(x,xd) -x + f(M*xd + [P*ones(N,1); Q*ones(N,1)]);
\end{lstlisting}
Here, the current state \texttt{x} and the delayed state \texttt{xd} each hold the $N$ excitatory activities followed by the $N$ inhibitory ones, and $M$ is a $2N\times2N$ block matrix.

The calculation of \Cref{sec:WCHopf} carries over when all the delays are equal and every region has the same equilibrium activity $(u^*,v^*)$, as happens in symmetric networks such as a ring. Each eigenvalue $\gamma$ of the weight matrix $W=\left(w_{ij}\right)$ then gives one copy of the stability calculation of \Cref{sec:WCHopf}, at the same equilibrium, with $a$ replaced by $a+\varepsilon\gamma$ (see \Cref{rem:nonsym}). Each pattern of activity across the network (each eigenvector of $W$) therefore has its own sequence of Hopf delays (\ref{eqn:HopfDelays}). Our in-phase and anti-phase modes are the simplest example of this idea, with the eigenvectors $(1,1)^T$ and $(1,-1)^T$ of \Cref{sec:WCHopf} playing the role of the network patterns. The pattern that loses stability first determines which regions oscillate together, which is how delays shape synchrony in brain networks
\cite{al2021impact,pinder2020oscillations,pinder2023bifurcations,pinder2024network}.

In this setting, simulation remains easy but continuation becomes expensive: DDE-BIFTOOL stores all $2N$ activities at every point along the orbit, so the systems it must solve grow quickly with $N$. For more than a few tens of regions, simulation is usually the main tool. Alternatively, in the limit of weak coupling between populations, key features of the solutions available may be determined analytically by weakly coupled oscillator theory, as described by \cite{Nelson2026phase}.

\subsection{Other types of delay}
\label{sec:delaytypes}
Real transmission delays are neither single nor exactly constant: different pathways have different lengths, and fibres in a bundle conduct at different speeds.

If, for example, each population feels its own activity with one (constant) delay and the other populations' with another, the equations simply contain both delayed states. DDE-BIFTOOL can handle this directly: \texttt{sys\_tau} returns the positions of both delays in the parameter vector (for example \texttt{@() [8 9]}), and the right-hand side
then receives the current state in \texttt{x(:,1)} and the two delayed states in \texttt{x(:,2)} and \texttt{x(:,3)}. The Hopf delays can no longer, in general, be found by hand, so continuation in two parameters, as in \Cref{fig:5}, becomes the natural way to explore the model. More generally, delays may depend on distance between brain regions or upon the state of the system itself. For state-dependent delays, we can replace our use of MATLAB's \texttt{dde23} above with the companion function \texttt{ddesd}, and can modify the corresponding DDE-BIFTOOL codes to account for state dependence as described by \cite{sieber2014ddebiftool}.

\subsection{Noise}
\label{sec:noise}
Real neural populations are noisy, because they contain a finite number of neurons and receive inputs that the model does not describe. The simplest approach to incorporating noise adds a small random kick at every step of the Euler map (\ref{eqn:EulerMap}):
\[
  \mathbf{x}_{n+1}=\mathbf{x}_n+h\,\mathbf{F}(\mathbf{x}_n,\mathbf{x}_{n-m})
    +\sigma\sqrt{h}\,\boldsymbol{\xi}_n,
\]
where each $\boldsymbol{\xi}_n$ is a vector of independent random numbers from the standard normal distribution and $\sigma$ sets the strength of the noise. This is the Euler--Maruyama method \cite{buckwar2000introduction}. In Listing~\ref{lst:eulermap} it amounts to adding \texttt{+ sigma*sqrt(h)*randn(2,1)} to the update. In the limit of small noise, we would anticipate the corresponding bifurcation analysis to approximately recover the results above; however, due to the additional instability that the noise represents, irregular oscillations and complex solutions generally appear in simulations slightly before the predicted bifurcation values in the above analysis. We note also that the addition of noise renders quasi-periodic, mode-locked and weakly chaotic motion harder to tell apart. The bifurcation diagrams of the model without noise remain the best guide to what the noisy model will do \cite{bressloff2010stochastic}.

\subsection{Applications in neuroscience}
\label{sec:applications}
Delayed neural mass models are widely used to study brain rhythms and their disorders. For example, in Parkinson's disease, patients show excessive oscillations at 13--30 oscillations per second (the beta band) in a loop between two interconnected regions deep in the brain, the subthalamic nucleus and the globus pallidus. Models of this loop of Wilson--Cowan type, with a delay on each connection, identify the delays and connection strengths for which the beta oscillations appear \cite{pavlides2015computational}; the analysis is a version of our calculation of the Hopf delays above. Similarly, in epilepsy, neural mass models reproduce features of the brain activity recorded at the start of seizures \cite{wendling2002epileptic}, and the way seizures start and stop has been classified by the bifurcations through which a model enters and leaves its seizure state \cite{jirsa2014nature}. The same tools apply to the rhythms recorded by EEG and MEG \cite{david2003neural,jansen1995electroencephalogram}. In each case, the questions are the ones we have asked here: at what delay does the resting state become unstable, what replaces it, and how does that change as the parameters vary?

\section{Discussion and Further Reading}
\label{sec:conclusions}

Using the Wilson--Cowan model with a single delay, we have introduced the main analytical and numerical tools for studying delay differential equations. The model is simple enough for its first bifurcation to be found by hand, yet rich enough to display torus bifurcations, period doubling, isolas of periodic orbits and chaos as the delay alone is varied.

We have seen that the delay does not change the equilibria of the model, only their stability (\Cref{sec:WCHopf}). For the symmetric model, the characteristic equation factorises into in-phase and anti-phase parts, and the delays at which the equilibrium loses stability can be written down exactly: they form an infinite, equally spaced sequence of Hopf bifurcations, and at every one of them the equilibrium loses stability. Direct simulation with the Euler map and with \texttt{dde23} confirmed this picture (\Cref{sec:NumSims}). Numerical continuation with DDE-BIFTOOL then allowed
us to follow equilibria and periodic orbits as the delay varies, to compute their stability through eigenvalues and Floquet multipliers, and to locate torus bifurcations, period doublings and folds of periodic orbits
(\Cref{sec:CompBifn}). We also observed that not every orbit of interest is born at a Hopf bifurcation: the asymmetric orbits of \Cref{fig:2}(b) lie on an isola, which can only be reached from a good initial guess, here a simulation. Two-parameter continuation (\Cref{fig:5}) placed these one-parameter results in a wider context. Finally, simulations and Lyapunov exponents showed how the torus bifurcation of the anti-phase orbit leads, through quasi-periodic motion and
mode locking, to chaos (\Cref{sec:chaos}).

Our aim throughout this tutorial has been to provide an accessible introduction to the topic of delay differential equations and their simulation and analysis. For the reader interested in finding out more, we recommend the accessible introductions of \cite{erneux2009applied} and \cite{smith2011introduction}, with more advanced treatments in \cite{diekmann2012delay} and \cite{hale2013introduction}. Numerical methods for solving DDEs are covered in \cite{bellen2003numerical} and \cite{shampine2001solving}, and the numerical computation of the stability of equilibria of DDEs is covered in more detail in
\cite{breda2015stability}. For bifurcation theory in general, including the torus, period-doubling and fold bifurcations met here, see \cite{kuznetsov2004elements,strogatz2015nonlinear}. In order to continue development of more advanced understanding of DDE-BIFTOOL code beyond that shown in the listings in this paper and the supporting codes online, we refer the reader to the DDE-BIFTOOL manual \cite{sieber2014ddebiftool} and the excellent demonstrations distributed with the package itself. For those preferring alternatives to Matlab, the Julia library \texttt{DDEBifurcationKit.jl} offers similar capabilities to DDE-BIFTOOL \cite{veltz_ddebifurcationkit}. Similarly, MatCont, a widely used MATLAB package for the continuation of ODEs with a graphical interface, has also recently been extended to delay equations and to the computation of Lyapunov exponents \cite{liessi2026new}, providing an alternative to DDE-BIFTOOL within MATLAB. The role of delays in neural systems is reviewed in \cite{campbell2007time}, and the delayed Wilson--Cowan model studied here was analysed in detail in \cite{coombes2009delays}. Wider introductions to mathematical neuroscience are given by \cite{dayan2005theoretical} and \cite{ermentrout2010mathematical}, and large-scale brain network models are reviewed in \cite{breakspear2017dynamic} and \cite{deco2008dynamic}. 

In conclusion, time delays are an unavoidable feature of neural communication, and even a small delay can significantly change the qualitative behaviour of a neural population model. Because the effects of a delay cannot, in general, be guessed from the undelayed model, they must be computed. We hope that the combination of exact calculation, simulation and numerical continuation presented here, together with the accompanying code, will help readers to explore delayed models of their own.

\section*{Code availability}
The MATLAB and Python code accompanying this paper is available at \url{https://github.com/martinrnelson/WC_DDE}. The continuation codes require DDE-BIFTOOL version 3.x, which is freely available. The README file in the repository lists which script produces each figure.

\bibliographystyle{siamplain}
\bibliography{references}
\end{document}